\RequirePackage{fix-cm}
\documentclass[11pt,twoside,twocolumn,english]{article}
\usepackage{beraserif}
\usepackage{biolinum}

\usepackage[T1]{fontenc}
\usepackage[latin9]{inputenc}
\usepackage{geometry}
\usepackage{amsmath}
\usepackage{graphicx}
\usepackage{microtype}

\makeatletter
\newcommand{\lyxaddress}[1]{
	\par {\raggedright #1
	\vspace{1.4em}
	\noindent\par}
}

\@ifundefined{date}{}{\date{}}

\usepackage{cite}
\usepackage[below]{placeins}
\usepackage[super]{nth}
\usepackage[font={sf,small}, labelfont={bf}, indention=0cm, format=plain]{caption}

\let\originalleft\left
\let\originalright\right
\renewcommand{\left}{\mathopen{}\mathclose\bgroup\originalleft}
\renewcommand{\right}{\aftergroup\egroup\originalright}

\DeclareMathOperator\diag{diag}

\renewcommand\vec[1]{\ensuremath{\boldsymbol{#1}}}

\makeatother

\usepackage{babel}
\begin{document}
\title{Phase-sensitive interferometry of decorrelated speckle patterns}
\author{Dierck Hillmann\textsuperscript{1,2,$*$,$\dagger$}, Hendrik Spahr\textsuperscript{2,3,$\dagger$},
Clara Pfäffle\textsuperscript{2,3}, Sazan Burhan\textsuperscript{2},\\
Lisa Kutzner\textsuperscript{2,3}, Felix Hilge\textsuperscript{2,3},
and Gereon Hüttmann\textsuperscript{2,3,4}}
\maketitle

\lyxaddress{\textsuperscript{1}Thorlabs GmbH, Maria-Goeppert-Straße 9, 23562
Lübeck, Germany}

\lyxaddress{\textsuperscript{2}Institute of Biomedical Optics, University of
Lübeck, Peter-Monnik-Weg 4, 23562 Lübeck, Germany}

\lyxaddress{\textsuperscript{3}Medical Laser Centre Lübeck GmbH, Peter-Monnik-Weg
4, 23562 Lübeck, Germany}

\lyxaddress{\textsuperscript{4}Airway Research Center North (ARCN), Member of
the German Center for Lung Research (DZL), Germany}

\lyxaddress{\textsuperscript{{*}}dhillmann@thorlabs.com}

\lyxaddress{\textsuperscript{$\dagger$}These authors contributed equally}

\begin{abstract}
Phase-sensitive coherent imaging exploits changes in the phases of
backscattered light to observe tiny alterations of scattering structures
or variations of the refractive index. But moving scatterers or a
fluctuating refractive index decorrelate the phases and speckle patterns
in the images. It is generally believed that once the speckle pattern
has changed, the phases are scrambled and any meaningful phase difference
to the original pattern is removed. As a consequence, diffusion and
tissue motion below the resolution handicap phase-sensitive imaging
of biological specimen. Here, we show that, surprisingly, a phase
comparison between decorrelated speckle patterns is still possible
by utilizing a series of images acquired during decorrelation. The
resulting evaluation scheme is mathematically equivalent to methods
for astronomic imaging through the turbulent sky by speckle interferometry.
We thus adopt the idea of speckle interferometry to phase-sensitive
imaging in biological tissues and demonstrate its efficacy for simulated
data and for imaging of photoreceptor activity with phase-sensitive
optical coherence tomography. The described methods can be applied
to any imaging modality that uses phase values for interferometry. 
\end{abstract}

\section{Introduction}

Phase sensitive, interferometric imaging measures small changes in
the time-of-flight of a light wave by detecting changes of its phase.
But in many applications, statistical optical properties of the scattering
structure randomize the phase of the backscattered light, resulting
in a speckle pattern with random intensity and phase \cite{goodman2007speckle}.
As a consequence, we can only extract meaningful phase differences
from images with identical or at least almost identical speckle patterns
\cite{Creath:85}. However, if the detected wave's speckle pattern
changes over time, it inevitably impedes phase sensitive imaging \cite{Joenathan:99,Lehmann:97}.
For example, holographic interferometry or electronic speckle pattern
interferometry (ESPI) compare at least two states of backscattered
light acquired at different times, and it can only be applied if the
respective speckle patterns are still correlated \cite{Jones1977,Creath:85,Lehmann:97,Joenathan:99}.

Among the most important effects that decorrelate speckles with time,
are random motions and changes of the optical path length on a scale
below the resolution, e.g., by diffusion \cite{berne2000dynamic,doi:10.1119/1.19101}.
These effects are impossible to prevent. But also bulk sample motion
can cause the phase evaluation to compare different parts of the same
speckle pattern. While in 3D imaging this effect of bulk motion can
often be corrected by co-registration and suitable algorithms, in
2D sectional imaging we lack the data to correct this as the acquired
slices of the specimen change if the specimen moves perpendicularly
to the imaging plane. 

We recently demonstrated imaging of the activity of photoreceptors
and neurons by phase-sensitive full-field swept-source optical coherence
tomography (FF-SS-OCT) \cite{Hillmann2016,2018arXiv180902812P}. Although
FF-SS-OCT successfully measured nanometer changes of neurons and receptors
due to activation over a few seconds, we could neither increase the
measurement time, nor determine tissue changes in single cross-sectional
scans (B-scans). In both cases speckle patterns changed after a few
seconds due to diffusion, bulk tissue motion, or tissue deformations
and the phase information was lost. 

\begin{figure*}[!t]
\begin{centering}
\includegraphics[width=1\textwidth]{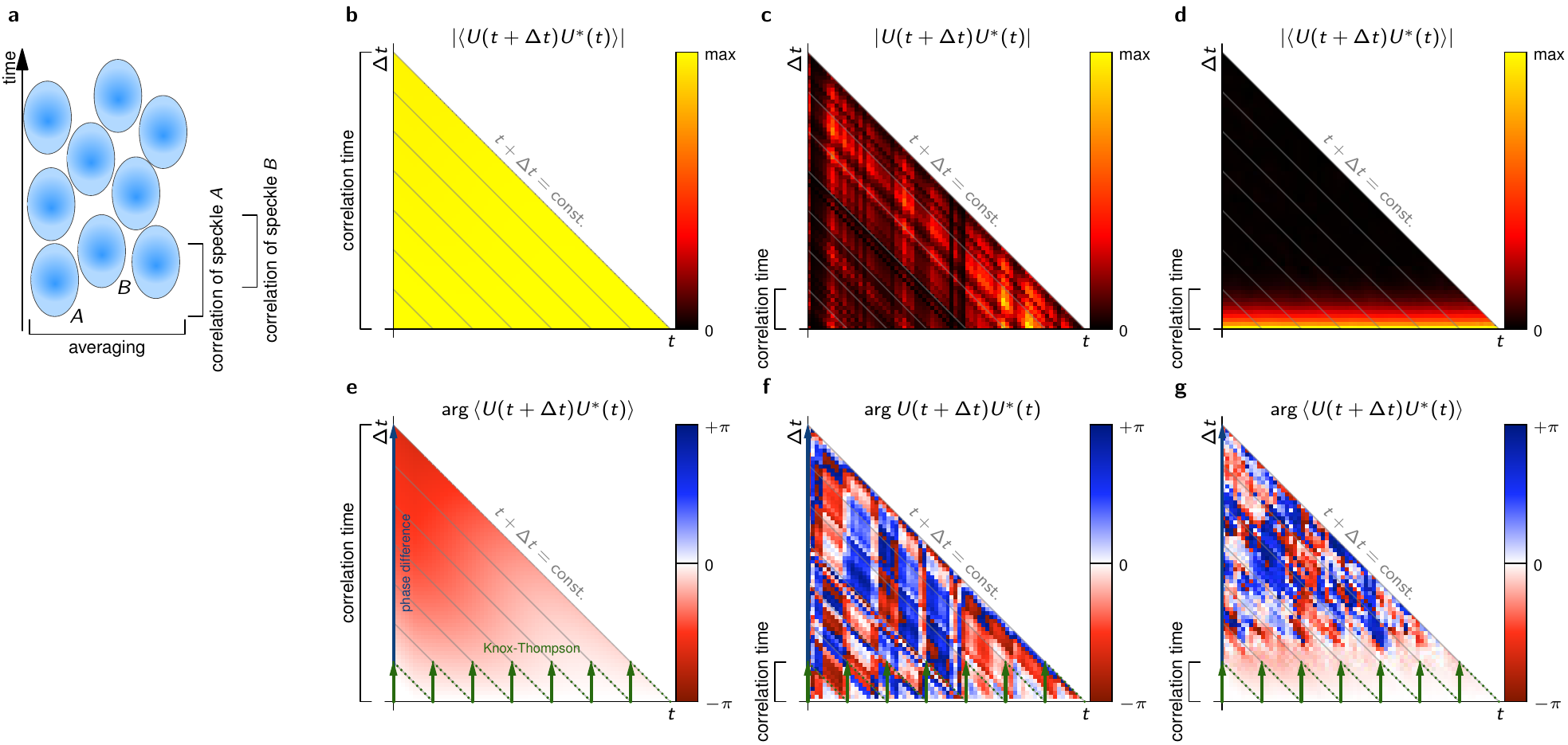}
\par\end{centering}
\caption{\label{fig:Schematic}a)~In average, each speckle (schematically
represented by the blue ellipses) carries valid phase information
only for its correlation time. Assuming that all speckle in the averaged
area are subject to a common phase change in addition to random uncorrelated
phase changes, one can use the phase of speckle $A$ as long as it
is valid and then continue with the phase of speckle $B$. Using multiple
speckle, phase information for times significantly exceeding the correlation
time can be extracted. b, e)~Exemplary cross-spectrum magnitude and
phase assuming infinite correlation time. Total phase changes can
either be computed directly (blue arrow), or using the Knox-Thompson
path (green arrows). c, f) Magnitude and phase of the cross-spectrum
without averaging; neither method can extract the phase beyond the
correlation time. d, g)~Magnitude and phase of ensemble averaged
cross-spectrum. Direct phase differences (blue) cannot be used for
phase extraction, but Knox-Thompson (green) method can be applied
since phase values for small $\Delta t$ are valid.}
\end{figure*}
In this paper, we reconstruct phase changes over times significantly
longer than the decorrelation time of the corresponding speckle patterns.
The idea is to calculate phase difference in a series of consecutive
images over small time differences that still show sufficient correlation.
If this is done for several measurements with different speckle patterns,
averaging of these independent measurements will cancel the contributions
of the disturbing phase. Integrating all these phase differences then
yields the real time evolution of the phase. By averaging the phase
in multiple speckles to obtain a single phase value, the real phase
change can be obtained beyond the correlation time of the speckles.
In essence, we combine the information from multiple speckles, each
of which carries information on phase changes over a certain time
(see Fig.~\ref{fig:Schematic}a). 

Just integrating the phase difference of successive frames followed
by a summation, as it is commonly used for phase unwrapping \cite{10.1007/978-3-642-57323-1_2,doi:10.1063/1.2724920,doi:10.1117/1.JBO.17.7.076026,An:13,Spahr:15},
accumulates the phase noise; except for phase unwrapping it is mathematically
equivalent to a direct phase comparison. Without averaging, calculating
short time phase differences will not yield any information about
phase differences of uncorrelated speckle patterns. The averaging
of short time phase differences before integration is the essential
step as it effectively cancels the phase noise.

In the 1970s and 1980s the same idea, known as speckle interferometry,
was originally used in astronomy for successful imaging through the
turbulent atmosphere with diffraction limited resolution \cite{Knox1974,Lohmann:83,Ayers:88,Lannes1989,Marron:90,Haniff:91,roggemann1996imaging,Mikurda2006,goodman2007speckle}.
Short time exposures were used to reconstruct the missing phase of
the Fourier transform of diffraction limited undisturbed images: By
utilizing a large number of short-exposure images with different disturbances,
true and undisturbed phase differences for small distances in the
aperture plane could be computed. From these phase differences the
complete phase was reconstructed by integration. We will frame the
similarities and the mathematical analogy between the two methods,
which allows us to benefit from advances in astronomy for interferometric
measurements during coherent imaging. 

Methods that allow phase evaluation beyond the correlation time have
also been developed for synthetic aperture radar (SAR) interferometry
(InSAR) \cite{Bamler1993,Zebker1992}. In InSAR, phase evaluation
is used to monitor ground elevation, but small changes on the ground
or turbulence in the sky interfere with an evaluation of the phase
\cite{Bamler1993} similar to interferometric measurements in biological
imaging. For InSAR, two techniques have been developed which selectively
evaluate the phase from only minimally affected images or structures.
The first technique is referred to as small baseline subset technique
(SBAS, \cite{10.1007/978-3-7643-8417-3_2}). It selects optimal images
from a time series to compare phase values based on good correlation.
The other method only compares phases of single selected scatterers
that maintain good correlation over long times, so called permanent
scatterers \cite{898661}. 

In the first part of this paper, we lay down the basics and establish
the commonality between coherent phase-sensitive imaging and astronomic
speckle interferometry. Afterwards, we demonstrate the efficacy of
the resulting methods by simulating simple phase resolved images in
backscattering geometry and evaluating these with the proposed approach.
Finally, we apply our method to in vivo data of phase-sensitive optical
coherence tomography. In a future paper, we will concentrate on details
of the algorithm, optimizing the time step for computing the phase
differences, and showing its advantages in different applications. 

\section{Theory}

\subsection{Mathematical formulation of the problem}

We assume coherent imaging of the physical system, in which a deterministic
change of the optical path length, e.g., swelling or shrinking of
cells, is superimposed by random changes. These may be caused by diffusion,
fluctuations of the refractive index, or rapid uncorrelated micro
motion. The complex amplitude in each pixel of a coherently acquired
and focused image can be represented as a sum of (random) phasors
(complex values), where each summand comprises amplitude and phase.
Such a coherently acquired field $U\left(\vec{x},t\right)$ has contributions
from the random phases $\phi_{i}\left(\vec{x},t\right)$ and a systematic
phase $\phi\left(\vec{x},t\right)$. $U\left(\vec{x},t\right)$ may
then be written as

\begin{align}
U\left(\vec{x},t\right) & =\sum_{\mathrm{i}}A_{i}\left(\vec{x},t\right)\mathrm{e}^{\mathrm{i}\phi_{i}\left(\vec{x},t\right)}\mathrm{e}^{\mathrm{i}\phi\left(\vec{x},t\right)}\text{,}\label{eq:RandomSystematicPhase}
\end{align}
where $A_{i}$ represents the unknown amplitude of each scatterer,
and $\vec{x}$ and $t$ denote location and time, respectively. Depending
on the imaging scenario, $U$ might not depend on the location $\vec{x}$
at all or $\vec{x}$ might be up to three-dimensional as in tomographic
imaging. 

We can now separate $U$ into its systematic phase contribution $U_{0}$
and a random modifier $H$, which yields
\begin{align}
U(\vec{x},t) & =\underbrace{\left(\sum_{\mathrm{i}}A_{i}\left(\vec{x},t\right)\mathrm{e}^{\mathrm{i}\phi_{i}\left(\vec{x},t\right)}\right)}_{H\left(\vec{x},t\right)}\underbrace{\mathrm{e}^{\mathrm{i}\phi\left(\vec{x},t\right)}}_{U_{0}\left(\vec{x},t\right)}\label{eq:Modulation}\\
 & =H\left(\vec{x},t\right)U_{0}\left(\vec{x},t\right)\text{.}\nonumber 
\end{align}
The modifier $H$ covers all effects that alter the speckle patterns
over time, e.g., changing phase, changing amplitude, and scatterers
moving out of or into the detection area. In this form, we have no
way to distinguish $H$ from $U_{0}$. We can, however, compute the
ensemble average by 
\begin{equation}
\left\langle U\left(\vec{x},t\right)\right\rangle =\left\langle H\left(\vec{x},t\right)\right\rangle U_{0}\left(\vec{x},t\right)\text{,}\label{eq:PhaseModulBase}
\end{equation}
when assuming that $U_{0}$ (though not $H$) is constant over the
averaged ensemble. The ensemble could consist of repeated measurements
or certain pixels from a volume, in which $U_{0}$ is constant. In
this averaged expression, we can now obtain the time evolution of
$U_{0}$ for times where $\left\langle H\right\rangle $ remains approximately
constant, i.e., where the contribution of the random phases is small.
For $t$ exceeding the correlation time of $\left\langle H\right\rangle $,
however, the decorrelation of $\left\langle H\right\rangle $ still
prevents us from computing changes in the systematic phase $U_{0}$. 

\subsection{Solving the phase problem in astronomic speckle interferometry}

The resolution of large ground-based telescopes is limited by atmospheric
turbulences, which modify the phase of the optical transfer function
(OTF) and cause time varying speckles in the image. During a long
exposure these speckle patterns average out to a blurred image. Speckle
interferometry is a way to obtain one diffraction limited image from
several short exposure images. In each of these images the random
atmospheric turbulences are constant and each image essentially contains
one speckle pattern \cite{goodman2007speckle}. If the actual diffraction
limited image of the object is $I_{0}\left(\vec{x}\right)$ and has
a Fourier spectrum $\tilde{I}_{0}\left(\vec{k}\right)$, the Fourier
spectra of the short exposure images are given by 
\begin{equation}
\tilde{I}\left(\vec{k}\right)=\mathcal{H}_{S}\left(\vec{k}\right)\tilde{I}_{0}\left(\vec{k}\right),\label{eq:ShortTimeExposure}
\end{equation}
where $\mathcal{H}_{S}$ is the short exposure OTF \cite{goodman2007speckle}.
$\mathcal{H}_{S}$ represents the disturbing effect of the atmosphere,
changes randomly in each acquired image, and causes speckle noise
in each image $I\left(\vec{x}\right)$. A single short exposure image
$I\left(\vec{x}\right)$ will thus not yield the object information
because phase and magnitude of $\mathcal{H}_{S}$ are not known. However,
the magnitude of the spectrum $\tilde{I}_{0}\left(\vec{k}\right)$
of the diffraction limited images is obtained by averaging the squared
magnitude of all short exposure spectra $\tilde{I}\left(\vec{k}\right)$,
\begin{equation}
\left\langle \left|\tilde{I}\left(\vec{k}\right)\right|^{2}\right\rangle =\left\langle \left|\mathcal{H}_{S}\left(\vec{k}\right)\right|^{2}\right\rangle \left|\tilde{I}_{0}\left(\vec{k}\right)\right|^{2}\text{,}\label{eq:AstronomyMagnitudeSpectrum}
\end{equation}
since the squared short exposure OTF averaged over all measurements
$\left\langle \left|\mathcal{H}_{S}\left(\vec{k}\right)\right|^{2}\right\rangle $
is larger than zero and can be computed on the basis of known statistical
properties of $\mathcal{H}_{S}$ \cite{goodman2007speckle}. But the
magnitude $\left|\tilde{I}_{0}\right|$ of the Fourier spectrum only
allows computation of the autocorrelation of $I_{0}$. To obtain the
full diffraction limited image the phase of $\tilde{I}_{0}\left(\vec{k}\right)$
is required too. This phase is contained in the Fourier transforms
$\tilde{I}\left(\vec{k}\right)$ of the short time exposures, but
it is corrupted by a random phase contribution of the atmospheric
turbulences in the same way as the systematic phase in coherent imaging
is corrupted by random phase noise. 

Hence, retrieving diffraction limited images by speckle interferometry
faces the same problem as retrieving the temporal phase evolution
in coherent imaging. Only over small spatial frequency distances in
the Fourier plane the phase difference in a spectrum $\tilde{I}\left(\vec{k}\right)$
corresponds to the phase difference in the spectrum $\tilde{I}_{0}\left(\vec{k}\right)$
of the diffraction limited image. Over larger distances the phase
information is lost due to atmospheric turbulence. The similarity
of the problem is seen in the analogy of Equations \eqref{eq:ShortTimeExposure}
with \eqref{eq:Modulation}, where $\mathcal{H_{S}}$ and $I$ correspond
to $H$ and $U$, respectively. However, while the astronomic problem
has a two dimensional vector $\vec{k}$ of spatial frequencies as
dependent variable, the phase-sensitive imaging problem has only the
time $t$. Both use a series of measurements, in which only phase
information over small $\Delta\vec{k}$ or $\Delta t$ is contained,
respectively. We will show that by analyzing a large number of measurements,
in which the corrupting phase contribution changes, the full phase
can be reconstructed. 

\subsection{Phase information in the cross-spectrum}

In astronomy, it were Knox and Thompson who solved the recovery of
phases from statistical short exposure images \cite{Knox1974} using
the so-called cross-spectrum. Later, another approach, called the
tripple-correlation or bispectrum technique, even improved the achieved
results \cite{Lohmann:83}. One major advantage of the bispectrum
over the cross-spectrum technique for astronomic imaging is that it
is only sensitive to the non-linear phase contributions to the transfer
function $\mathcal{H}_{S}$, since shifted images yield to the same
bispectrum and do not pollute the obtained phases (see, for example,
\cite{Ayers:88} for details). However, for phase sensitive imaging,
the linear part of the phase evolution of $U_{0}$ is generally of
interest and thus the bispectrum technique is not applicable here. 

Following we will apply the algorithm of Knox and Thompson to recover
the phase beyond the speckle correlation time in coherent imaging.
For convenience, we will retain the term cross-spectrum, even though
in our scenario it is evaluated in time-domain as a function of $t$
rather than in (spatial) frequency-domain as a function of $\vec{k}$. 

We define the cross-spectrum $C_{H}$ of the phase disturbing speckle
modulation $H$ by 
\[
C_{H}\left(\vec{x},t,\Delta t\right)=H\left(\vec{x},t\right)H^{*}\left(\vec{x},t+\Delta t\right)\text{.}
\]
It is subject to speckle noise since $H$ itself merely contains speckle
(Fig.~\ref{fig:Schematic}c and f); but its ensemble average
\begin{equation}
\left\langle C_{H}\left(\vec{x},t,\Delta t\right)\right\rangle =\left\langle H\left(\vec{x},t\right)H^{*}\left(\vec{x},t+\Delta t\right)\right\rangle \label{eq:ModulationAvgCrossSpectrum}
\end{equation}
has in general a magnitude larger than $0$ and speckles of $H$ are
averaged out, at least for small $\Delta t$. Most importantly, it
is to a good approximation real-valued, i.e., its phase is zero, as
long as $\Delta t$ is within the correlation time of $H$. This real-valuedness
was previously shown for $\left\langle C_{\mathcal{H}_{S}}\right\rangle $
in astronomic imaging \cite{roggemann1996imaging,goodman2007speckle}.
For phase imaging, we show the real-valuedness of $\left\langle C_{H}\right\rangle $
for small $\Delta t$ for which the autocorrelation is still strong
in the Methods section. If $\Delta t$ is larger than the correlation
time, $H\left(\vec{x},t\right)$ and $H^{*}\left(\vec{x},t+\Delta t\right)$
become statistically independent, the phase of $\left\langle C_{H}\right\rangle $
gets scrambled and $C_{H}$ follows speckle statistics. 

Now, if we again assume that $U_{0}$ and thus $C_{U_{0}}$ are constant
over the ensemble, we compute the averaged cross-spectrum of the measurements
of $U$ to
\begin{multline*}
\left\langle C_{U}(\vec{x},t,\Delta t)\right\rangle =\left\langle U\left(\vec{x},t\right)U^{*}\left(\vec{x},t+\Delta t\right)\right\rangle \\
=\left\langle C_{H}\left(\vec{x},t,\Delta t\right)\right\rangle C_{U_{0}}\left(\vec{x},t,\Delta t\right)\text{.}
\end{multline*}
Knowing that $\left\langle C_{H}\right\rangle $ is real-valued for
small $\Delta t$, the phases of $\left\langle C_{U}\right\rangle $
are determined only by the phases of $C_{U_{0}}$. If we find phases
of $U_{0}$ that yield the phases of $\left\langle C_{U}\right\rangle $
for these small $\Delta t$, we obtain the systematic phase $\phi\left(\vec{x},t\right)$
we are looking for as introduced in \eqref{eq:RandomSystematicPhase}.

Note that $\left\langle C_{U}\left(\vec{x},t,\Delta t\right)\right\rangle $
is related to the time-autocorrelation of $U$, except averaging being
performed over the ensemble instead of the time $t$ and thus being
a function not only of $\Delta t$ bus also of $t$. Due to this relation
the cross-spectrum maintains comparably large magnitudes for time
differences $\Delta t$ with large autocorrelation values. The magnitudes
and phases of exemplary cross-spectra, in the decorrelation-free scenario
and with strong decorrelation, with and without the ensemble averaging
(obtained from simulations, see Results and Methods) is shown in Figs.~\ref{fig:Schematic}b
to g. The figures illustrate that $\left\langle C_{U}\right\rangle $
yields a deterministic increase of the phases for small $\Delta t$
(Fig.~\ref{fig:Schematic}g) as long as its magnitude (Fig.~\ref{fig:Schematic}d,
related to the autocorrelation) remains large, but only if the ensemble
averaging is performed. It can further be seen, that the valid phase
differences visible in Fig.~\ref{fig:Schematic}g are small compared
to the phases for large $t$ in Fig.~\ref{fig:Schematic}e since
they represent merely phase difference for small $\Delta t$. Nevertheless,
when reconstructing the phase of $U_{0}$ from the values of $\left\langle C_{U}\right\rangle $,
we need to ensure that only values for (small) $\Delta t$ with strong
autocorrelation are taken into account. This condition coincides with
the real-valuedness of $\left\langle C_{H}\right\rangle $.

Ensemble averaging needs multiple independent measurements, which
cannot be performed easily in a real experiment. However, averaging
over multiple lateral pixels or averaging over multiple detection
apertures can be done from a single experiment and approximates the
ensemble average. In our case we either use an area over which we
strive to obtain one mean phase-curve, or we use a Gaussian filter
with a width determined as a compromise between spatial resolution
and sufficient averaging statistics to obtain images of the phase
evolution (see Methods).

\subsubsection{Phase retrieval from the cross-spectrum}

Having obtained the cross-spectrum $\left\langle C_{U}\right\rangle $,
the systematic phase function $U_{0}$ needs to be extracted. We assume
a given initial phase $\phi(\vec{x},t=t_{0})$ and evaluate methods
to extract the systematic phase evolution $\phi\left(\vec{x},t\right)$
from the cross-spectrum: An approach equivalent to computing phase
differences directly is given by 
\begin{equation}
\phi(\vec{x},t_{0}+t)=\phi\left(\vec{x},t_{0}\right)+\arg\left\langle C_{U}(\vec{x},t_{0},t)\right\rangle \text{,}\label{eq:PhaseDifferences}
\end{equation}
for $t\ge t_{0}$ (blue arrows in Figs.~\ref{fig:Schematic}b-g).
However, it only yields phases within the correlation time. In astronomy,
this approach never works, unless the image was not disturbed by turbulence
in the first place. Instead, Knox and Thompson originally demonstrated
diffraction limited imaging through the turbulent sky\cite{Knox1974}
by iteratively walking in small increments of $\Delta t$ (e.g., one
time step) through the cross-spectrum, i.e., 
\begin{multline}
\phi(\vec{x},t_{0}+n\Delta t)=\phi(\vec{x},t_{0}+(n-1)\Delta t)\\
+\arg\left\langle C_{U}(\vec{x},t_{0}+(n-1)\Delta t,\Delta t)\right\rangle \text{,}\label{eq:IntPhaseDifferences}
\end{multline}
for integer $n\ge1$ with increasing iteration number $n$ (green
arrows in Figs.~\ref{fig:Schematic}b-g). This method uses only phase
values well within the correlation time. If $\left\langle C_{U}\right\rangle $
is valid for the single time steps $\Delta t$, i.e., $\left\langle C_{H}\right\rangle $
is real valued, the iterative formula will yield valid results, even
for larger $t$. However, a single outlier at a time $t_{\text{err}}$,
i.e., one false step, ruins all following values for $t\ge t_{\text{err}}$. 

\begin{figure*}
\centering{}\includegraphics[scale=0.8]{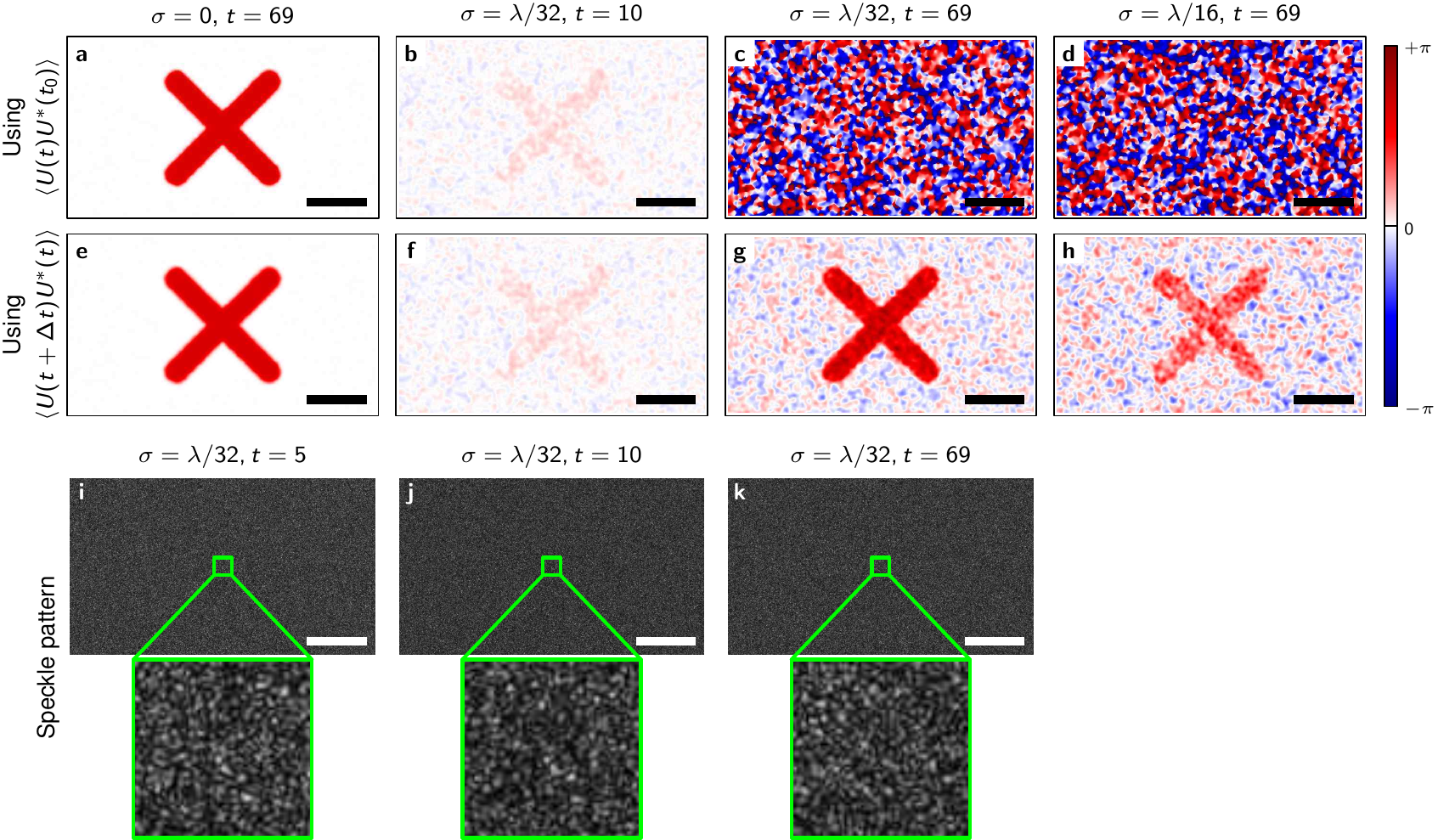}\caption{\label{fig:SimImages}Phase images and speckle patterns obtained from
the simulation. a-d)~Phase evaluated by phase differences for different
$\sigma$ and time steps. e-h)~Phase evaluated using the extended
Knox-Thompson methods for the same $\sigma$ and time steps as shown
in (a)-(d). i-k)~Speckle patterns for $\sigma=\lambda/32$ after
a different number of time steps. Scale bars are $500\lambda$.}
\end{figure*}
The influence of these single events can be reduced by taking multiple
ways with different $\Delta t$ for stepping through the cross-spectrum.
In astronomy, this is known as the extended Knox-Thompson method \cite{Ayers:88,Mikurda2006}.
However, most of the algorithms for finding the optimal paths have
been developed for use with respect to the bispectrum technique \cite{Lannes1989,Marron:90,Haniff:91};
nevertheless, in general they can be easily transferred to the cross-spectrum.
Here we minimized the sum of weighted squared differences between
the measured cross-spectrum and the cross-spectrum resulting from
the systematic phase $\phi\left(\vec{x},t\right)$. The algorithm
is described step-by-step in the methods section, and we will motivate,
describe, and evaluate it in more depth in a future publication.

\section{Results}

\subsection{Simulation}

We simulated images of a large number of point scatterers that exhibit
a random Gaussian-distributed 3-dimensional motion (with variance
$\sigma^{2}$) in between frames in addition to a common axial motion
in a specific ``x''-shaped area. This simulation of degrading speckle
clearly demonstrates the power of the extended Knox-Thompson method.
For $\sigma=\lambda/32$, the phase evaluation with a simple phase
difference to the first image and the extended Knox-Thompson method
yield almost indistinguishable results for the first $10$ steps (Fig.~\ref{fig:SimImages}b
and f). After $60$ time steps, a simple phase difference to the first
image yields only noise (Fig.~\ref{fig:SimImages}c), while the ``x''-shaped
pattern is still well visible (Fig.~\ref{fig:SimImages}g) using
the extended Knox-Thompson method. The speckle patterns appear correlated
after $10$ steps, but are completely changed after $69$ steps (Figs.~\ref{fig:SimImages}i-k).
Doubling the scatterer motion to $\sigma=\lambda/16$, the ``x''-shaped
pattern begins to deteriorate after 69 steps even with the Knox-Thompson
method. However, the ``x'' remains visible (Fig.~\ref{fig:SimImages}h),
whereas the phase difference again yields only phase noise (Fig.~\ref{fig:SimImages}d).

\begin{figure*}[t]
\centering{}\includegraphics[width=1\textwidth]{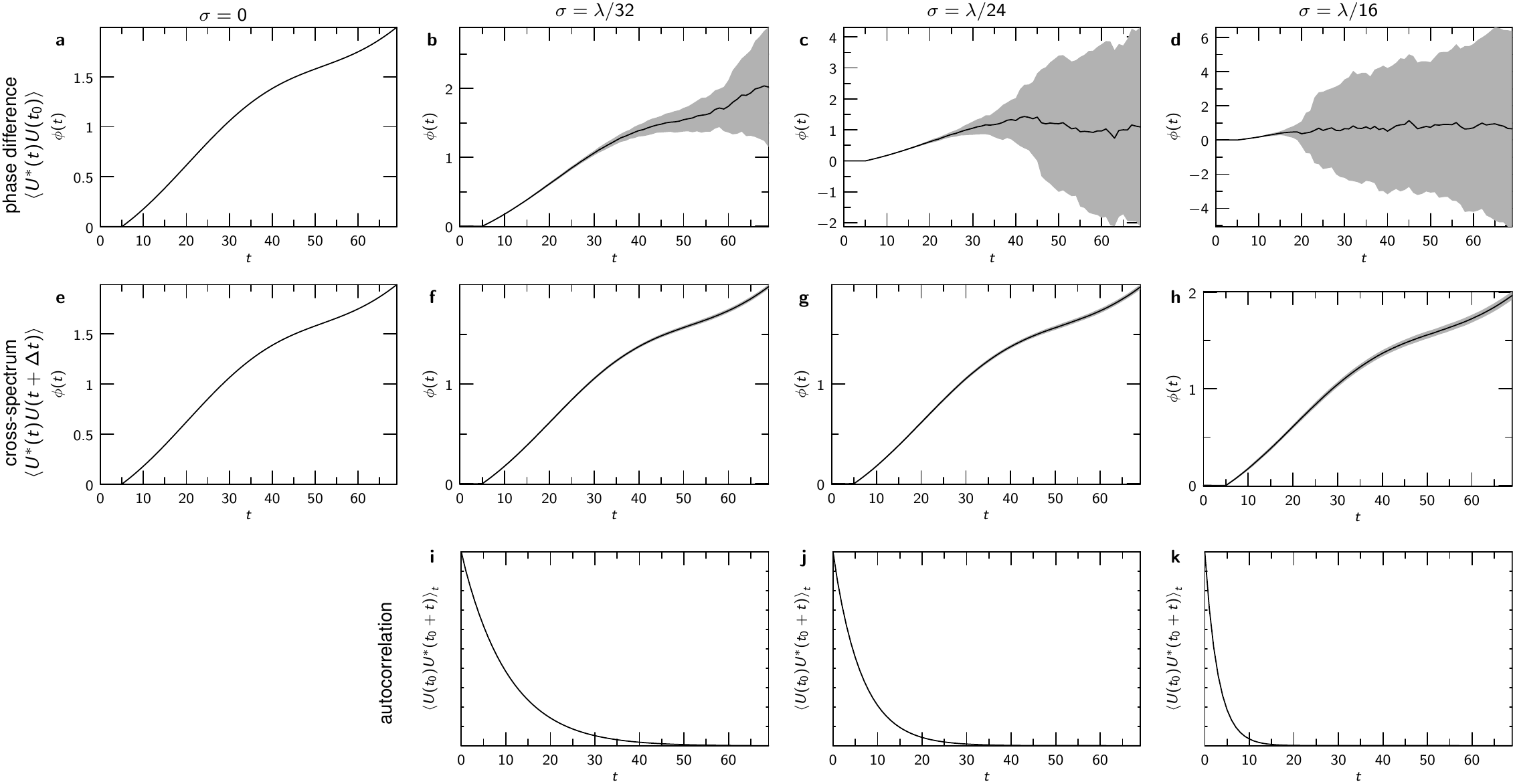}\caption{\label{fig:SimCurves}Extracting a single phase curve from simulated
images of point scatterers that move between successive measurements
by subtracting the phase of the first image (a-d) and Knox-Thompson
evaluation of phases (e-h) while random 3D motion of the simulated
scatterers is increased from 0 to $\lambda/16$. i-k)~Autocorrelation
of temporal changes of the wave field. The autocorrelation has been
normalized to $1$ for $t=0$. Each curve was simulated $100\times$.;
the gray areas indicate the standard deviation of the obtained values. }
\end{figure*}
The curves extracted in multiple simulations from the ``x'' confirm
the results of the images and demonstrate the reproducibility of the
method. Without random scatterer motion ($\sigma=0$, Fig.~\ref{fig:SimCurves}a,e
and Fig.~\ref{fig:SimImages}a,e), the phase can be calculated by
the phase difference to the first frame yielding the exact curve that
was supplied to the simulation. But introducing random Gaussian-distributed
motion of the scatterers with as little as $\sigma=\lambda/32$ between
frames accumulates huge errors in the phase of the directed motion
after 70 frames (Fig.~\ref{fig:SimCurves}b); the mean autocorrelation
of the complex fields (Fig.~\ref{fig:SimCurves}i) in this case drops
to one half in less than ten frames. Increasing motion amplitude (Fig.~\ref{fig:SimCurves}c
and d) has devastating effects on the calculated phase differences.
Retrieving the phase with the cross-spectrum based Knox-Thompson method
yields the directed motion up to $\sigma=\lambda/16$ (Fig.~\ref{fig:SimCurves}f,
g, and h) even though the autocorrelation halves after few frames
(Fig.~\ref{fig:SimCurves}k).

\subsection{In vivo experiments}

\begin{figure*}
\begin{centering}
\includegraphics[scale=0.9]{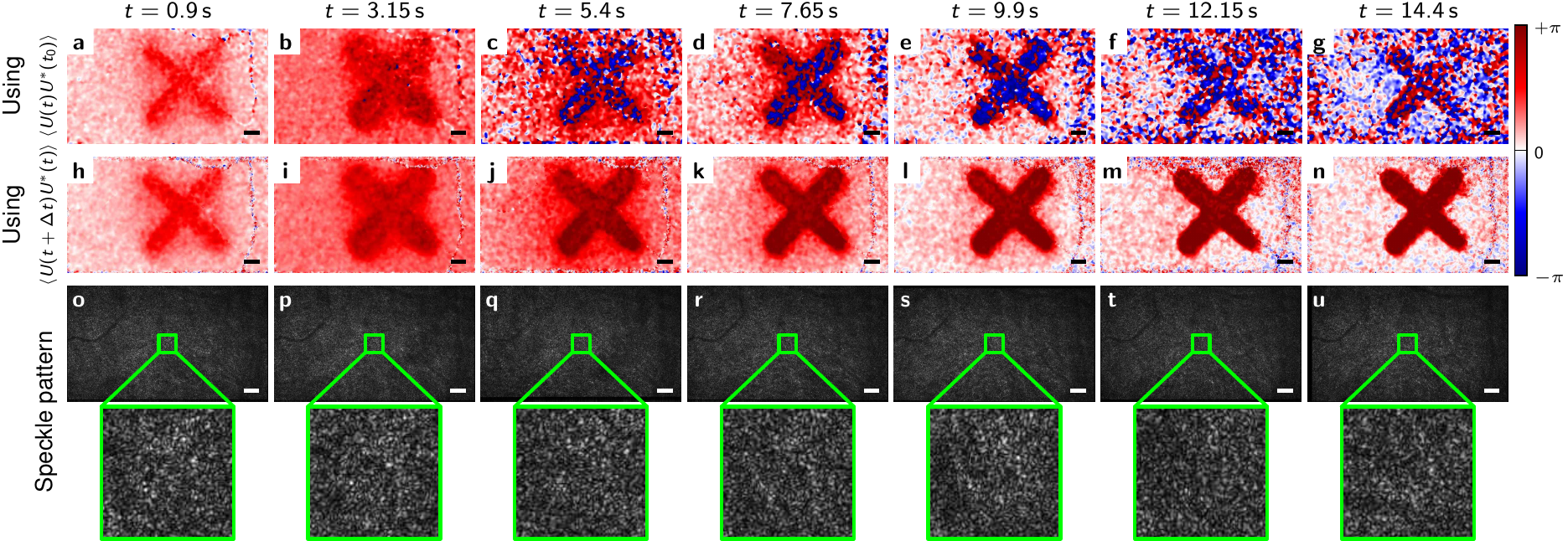}
\par\end{centering}
\caption{\label{fig:ExpImages}Phase differences between the ends of the photoreceptor
outer segment in the living human eye at different times compared
to the initial phase at $t=0$. The entire measurement ended at about
$t=14.6\,\mathrm{s}$ after initiating a light stimulus. a-g)~Phase
difference obtained by directly comparing to pre-stimulus phase with
phase after different times. h-n)~Phase difference obtained from
the cross-spectrum as described in the Methods section (by minimizing
\eqref{eq:CS-Minimization}). o-u)~Speckle pattern of one of the
involved layers before the stimulus and at the respective times. It
can clearly be seen, that that the speckle pattern changes with time.
Scale bars are $200\,\mathrm{\mu m}$.}
\end{figure*}
The in vivo experiments confirm the simulation results. Here, we imaged
human retina with full-field swept-source optical coherence tomography
(FF-SS-OCT, Fig.~\ref{fig:Setup}), which provides three-dimensional
tomographic data comprising amplitude and phase, to show the elongation
of photoreceptor cells in the living human retina. Phase differences
between both ends of the photoreceptor outer segments were evaluated
after stimulating the photoreceptors with an ``x''-shaped light-stimulus,
similar to the simulation (see Methods). Figs.~\ref{fig:ExpImages}a-g
show the extracted phase difference of the 5th frame (the beginning
of the stimulus) to various other frames of the data set. The last
frame was acquired 14.6 seconds later. After 5 seconds the phase differences
are dominated by random phases and the speckle patterns of the compared
images changed considerably after this time (Figs.~\ref{fig:ExpImages}o-u).
Diffusion and uncorrected tissue motion are likely to be the dominant
factors. The phase evaluation using the cross-spectrum is nevertheless
able to reconstruct the phase difference and show the elongation of
the photoreceptor outer segments (Figs.~\ref{fig:ExpImages}h-o).
Even in the non-stimulated area of Fig.~\ref{fig:ExpImages}g, it
is clearly seen that the phase difference gives random results, whereas
the extended Knox-Thompson approach shows the expected small changes
(Fig.~\ref{fig:ExpImages}n).

\section{Conclusion }

Building the phase differences of successive frames, and then summing
or integrating the differences has been used in phase sensitive imaging
\cite{10.1007/978-3-642-57323-1_2,doi:10.1063/1.2724920,doi:10.1117/1.JBO.17.7.076026,An:13,Spahr:15}.
It removes the burden of phase unwrapping and thereby improves results.
In contrast, including an average over multiple speckles after computing
the phase differences and before integrating those differences again,
as done by the Knox-Thompson method in astronomic speckle interferometry,
allows phase sensitive imaging over more than ten times the speckle
decorrelation time. To our knowledge, this important step has never
been fully realized in phase sensitive imaging. The resulting method
overcomes limitations from decorrelating speckle patterns and recovers
phase differences even if speckle are completely decorrelated. Using
extended Knox-Thompson methods yields the optimized results presented
here and removes the sensitivity to outliers.

Still, measurements and the algorithm can be further improved in several
regards. We approximated the ensemble average either by averaging
a certain area of lateral pixels or by a Gaussian filter. The former
is suitable to obtain the mean expansion in a certain area, the latter
is used to obtain images. But to get high-resolution images, using
spatial filtering is not ideal. Other speckle averaging techniques,
e.g., based on non-local means \cite{5617267,doi:10.1002/2013JE004584},
and time-encoded manipulation of the speckle pattern by deliberately
manipulating the sample irradiation (similar to \cite{Liba2017})
should preserve the full spatial resolution.

Increasing the sampling rate of the $t$ axis of the cross-spectrum,
i.e., decreasing the smallest $\Delta t$, should also improve results,
since then the correlation of compared speckle patterns is improved.
The results presented here are limited by the finite sampling interval
of the $t$-axis and not by the overall measurement time; however,
at some point it is expected that errors add up for high frequent
sampling of $t$ as data size increases further. Alternatively, one
could acquire two immediate frames with a small $\Delta t$, followed
by a larger time gap to the next two frames (during which speckle
patterns can begin to decorrelate). The immediate frames can give
the correct current rate of phase changes (the first derivative of
the phase change) that can then be extrapolated linearly to the time
of the next two frames, etc. Obviously, this can be extended to three
or more frames with small $\Delta t$ giving the second or even higher
order local derivative, respectively. 

There are also extensions of the presented method possible. For example,
the extraction of systematic phase changes in a dynamic, random speckle
pattern can be combined with dynamic light scattering (DLS)\cite{berne2000dynamic,doi:10.1119/1.19101}
basically separating the random diffusion from systematic particle
motion. This approach might possibly give additional contrast compared
to either method on its own. 

Overall, applications of phase sensitive imaging over long times are
manifold. Applications range from biological phase imaging to measure
retinal pulse waves \cite{Spahr:15}, detect cellular activity \cite{Hillmann2016,2018arXiv180902812P},
all the way to visualize deformations using electronic speckle interferometry. 

\bibliographystyle{naturemag}
\bibliography{PhaseEval}

\paragraph{Competing interests}

DH is an employee of Thorlabs GmbH, which produces and sells OCT systems.
DH and GH are listes as inventors on a related patent application.

\paragraph*{Acknowledgements}

This work was funded by the German Research Foundation (DFG), Project
Holo-OCT HU 629/6-1.

\section{{\small{}Materials and Methods}}

\subsection{{\small{}Simulation }}

{\small{}For the simulation we created a complex valued image series
with $640\times368$ pixels and a pixel spacing of $4\lambda$, where
$\lambda$ is the simulated light wavelength. To obtain an image series
we created a collection of $50\times640\times368=11,776,000$ point
scatterers each getting a random $x$, $y$, and $z$ coordinate,
as well as an amplitude $A$. While the $x$ and $y$ coordinate was
distributed entirely random in the image area, the $z$ coordinate
was restricted between $0$ and $10\lambda$ and the amplitude was
equally distributed between $0$ and $1$. To create an image we iterated
over all scatterers, and summed the values $A\exp(\mathrm{i}2kz)$
for the pixel corresponding to the scatterer's $x$ and $y$ coordinate,
where $z$ is the scatterer's $z$-value, and $k=2\pi/\lambda$. This
basically corresponds to a phase one would obtain in reflection geometry
when neglecting any possible defocus. Finally the image was laterally
filtered simulating a limited numerical aperture (NA). }{\small\par}

{\small{}For each simulation we created a series of $70$ images;
starting with the initial scatterers, to create the next frame, we
moved each scatterer randomly as specified by a Gaussian distribution
with a certain variance $\sigma^{2}$ in $x$, $y$, and $z$ direction.
Furthermore, all scatterers currently having $x$ and $y$-coordinates
as found in a pre-created ``x''-shaped mask were additionally subjected
to the movement 
\[
\Delta z=\begin{cases}
\frac{\lambda}{400}+\frac{\lambda}{800}\sin\left(0.1\left(t-t_{0}\right)\right) & t\ge t_{0}\\
0 & t<t_{0}
\end{cases}\text{,}
\]
where $t$ is the frame number and $t_{0}=5$ in each frame. }{\small\par}

\subsection{{\small{}Experiments}}

{\small{}}
\begin{figure}
\centering{}{\small{}\includegraphics{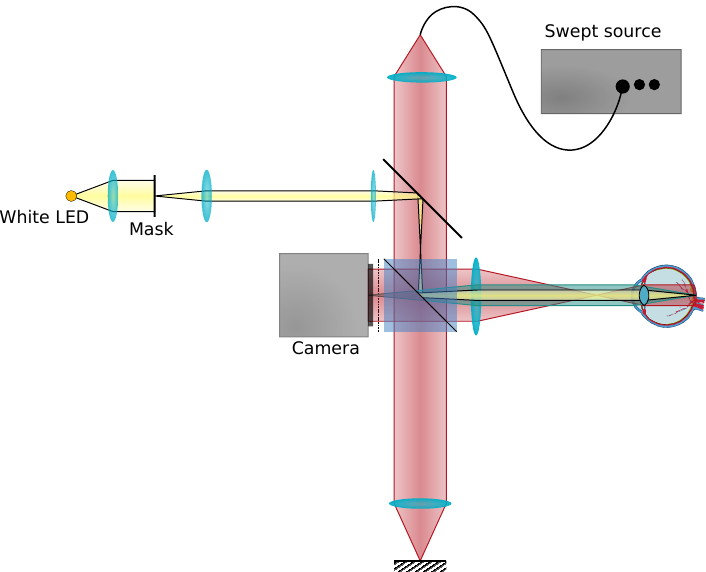}\caption{{\small{}\label{fig:Setup}Full-field swept-source optical coherence
tomography setup used for acquiring in vivo data.}}
}
\end{figure}
{\small{}In vivo data was acquired with a Michelson interferometer-based
full-field swept-source optical coherence tomography system (FF-SS-OCT)
as shown in Fig.~\ref{fig:Setup}. The setup is similar to the system
previously used for full-field and functional imaging \cite{Spahr:15,Hillmann2016,Hillmann2:2016}.
Light from a Superlum Broadsweeper BS-840-1 is collimated and split
into reference and sample arm. The reference arm light is reflected
from a mirror and then directed by a beam splitter onto a high-speed
area camera (Photron FASTCAM SA-Z). The sample light is directed in
such a way, that it illuminates the retina with a collimated beam;
the light backscattered by the retina is imaged through the illumination
optics and the beamsplitter onto the camera, where it is superimposed
with the reference light. }{\small\par}

{\small{}In addition the retina was stimulated with a white LED, where
a conjugated plane to the retina contained an ``x''-shaped mask.
A low-pass optical filter in front of the camera ensured that the
stimulation light does not reach the camera. }{\small\par}

{\small{}Swept laser, camera, and stimulation LED were synchronized
by an Ardunio Uno microprocessor board. It generates the trigger signals
such that the laser sweeps $70\times$ for one dataset. During each
sweep the camera acquires $512$ images with $640\times368$ pixels
each at a framerate of $60,000\,\mathrm{frames/s}$. Each set of 512
images corresponds to one OCT volume. The stimulation LED is triggered
after the first 5 volumes and remains active for the rest of the dataset.
The time between volumes determines the overall measurement time.
The number of $70$ volumes was limited by the camera's internal memory
since the acquisition of the camera is too fast to stream the data
to a computer in real time.}{\small\par}

{\small{}Measurement light had a centre wavelength of $841\,\mathrm{nm}$
and a bandwidth of $51\,\mathrm{nm}$. About $5\,\mathrm{mW}$ of
this light reached the retina illuminating an area of about $2.6\,\mathrm{mm}\times1.5\,\mathrm{mm}$,
which is the area imaged onto the camera. The ``x''-pattern stimulation
had significantly less irradiant power, of about $20\,\mathrm{\mu W}$
white light. }{\small\par}

\subsubsection{{\small{}In vivo data acquisition}}

{\small{}All investigations were done with healthy volunteers; written
informed consent was obtained from all subjects. Compliance with the
maximum permissible exposure (MPE) of the retina and all relevant
safety rules was confirmed by the responsible safety officer. The
study was approved by the ethics board of the University of Lübeck
(ethics approval Ethik-Kommission Lübeck 16-080).}{\small\par}

\subsubsection{{\small{}Data reconstruction}}

{\small{}Data was reconstructed, registered and segmented as described
in previous publications \cite{2018arXiv180902812P}: After a background
removal the OCT signal was reconstructed by a fast Fourier transform
(FFT) along the spectral axis. Next step was a dispersion correction
by multiplication in the complex-valued volumes in the axial Fourier
domain (after an FFT along the depth axis) with a correcting phase
function that was determined iteratively by an optimization of sharpness
metric of the OCT images \cite{Hillmann2:2016}. Co-registering the
volumes aligned the same structures in the same respective voxels
of the data series. Finally, segmenting the average volume allowed
aligning of the photoreceptor layers in a certain constant depth for
easier phase extraction. }{\small\par}

\subsection{{\small{}Phase evaluation}}

\subsubsection{{\small{}Computing the cross-spectrum}}

{\small{}To compute the cross-spectrum, we took the series of OCT
volumes $U(\vec{x},t)$ and computed the cross-spectrum by
\[
C_{U}(\vec{x},t,\Delta t)=U\left(\vec{x},t\right)U^{*}\left(\vec{x},t+\Delta t\right)\text{.}
\]
}{\small\par}

\subsubsection{{\small{}Comparing phases of two different depths}}

{\small{}For the in vivo experiments we additionally want to compare
the phases between different ranges of layers. Let $Z_{1}$ and $Z_{2}$
denote the sets of layers, then we computed the effective phase difference
cross-spectrum by 
\begin{multline}
C_{U}^{Z_{1}-Z_{2}}(x,y;t,\Delta t)=\left(\sum_{z\in Z_{1}}C_{U}(x,y,z;t,\Delta t\right)\\
\times\left(\sum_{z\in Z_{2}}C_{U}\left(x,y,z;t,\Delta t\right)\right)^{*}\text{,}\label{eq:CrossSpectrumPhaseDiff}
\end{multline}
where we now represented the vector $\vec{x}$ by its components $\left(x,y,z\right)$.
For the simulation, there is only one layer to be compared, and thus
we set the phase difference cross-spectrum equal to the cross-spectrum
of this one layer, i.e., $C_{U}^{Z_{1}-Z_{2}}=C_{U}$. }{\small\par}

\subsubsection{{\small{}Approximating the ensemble average}}

{\small{}To extract a curve the phase difference was approximated
by averaging the cross-spectrum over the $x$ and $y$ coordinates
belonging to a mask $M$, i.e.,
\[
\left\langle C_{U}\left(t,\Delta t\right)\right\rangle _{M}\approx\frac{1}{\left|M\right|}\sum_{\left(x,y\right)\in M}C_{U}^{Z_{1}-Z_{2}}\left(x,y;t,\Delta t\right)\text{,}
\]
where $\left|M\right|$ is the number of pixels in the mask $M$. }{\small\par}

{\small{}To obtain images, instead, the ensemble average was approximated
by applying a Gaussian filter (convolving with a Gaussian $G_{\sigma_{\mathrm{Gauss}}^{2}}$
with variance $\sigma_{\mathrm{Gauss}}^{2}$):
\begin{equation}
\left\langle C_{U}\left(x,y;t,\Delta t\right)\right\rangle \approx G_{\sigma_{\mathrm{Gauss}}^{2}}(x,y)*C_{U}^{Z_{1}-Z_{2}}\left(x,y;t,\Delta t\right)\text{,}\label{eq:GaussianFilter}
\end{equation}
where $*$ is the convolution in $x$ and $y$ direction. In effect
we used a circular convolution by using a fast Fourier transform (FFT)
based algorithm. }{\small\par}

\subsubsection{{\small{}The weights}}

{\small{}We additionally used weights to specify how reliable the
respective value of the cross-spectrum $\left\langle C_{U}\right\rangle $
is. Unfortunately, most algorithms that were used previously to compute
weights in astronomy cannot be computed as efficiently as $\left\langle C_{U}\right\rangle $,
since we used a fast convolution to apply the Gaussian filter in equation
\eqref{eq:GaussianFilter}. We therefore use the weights}{\small\par}

{\small{}
\begin{equation}
w(t,\Delta t)=T\left({\displaystyle \sum_{(x,y)\in M}\frac{C_{U}^{Z_{1}-Z_{2}}\left(t,\Delta t\right)}{\left|C_{U}^{Z_{1}-Z_{2}}\left(t,\Delta t\right)\right|}-b}\right)\label{eq:WeightsCurve}
\end{equation}
 for phase curve extraction and }{\small\par}

{\small{}
\begin{multline}
w(x,y;t,\Delta t)=T\Biggl(\Biggl[G_{\sigma_{\mathrm{Gauss}}^{2}}(x,y)\\
*\frac{C_{U}^{Z_{1}-Z_{2}}\left(x,y;t,\Delta t\right)}{\left|C_{U}^{Z_{1}-Z_{2}}\left(x,y;t,\Delta t\right)\right|}\Biggr]-b\Biggr)\label{eq:WeightsImaging}
\end{multline}
for phase imaging; both could be computed as efficiently as $\left\langle C_{U}\right\rangle $
itself. In these formula $T$ is a cut-off function for negative values:
\[
T\left(x\right)=\begin{cases}
x & \text{for \ensuremath{x\ge0}}\\
0 & \text{for }x<0
\end{cases}
\]
The truncation of negative values and the introduction of $b$ are
supposed to keep the expectation value for randomly distributed $C_{U}$
at $0$. The sum in \eqref{eq:WeightsCurve} and the convolution term
in \eqref{eq:WeightsImaging} follow speckle statistics if the phase
of $C_{U}$ is random, since they are a sum of random phasors \cite{goodman2007speckle}.
Consequently, a non-zero value is expected. This can be used to determine
suitable values for $b$. The amplitude probability of a speckle pattern
is given by a Rayleigh distribution \cite{goodman2007speckle} 
\[
p\left(A\right)=\frac{A}{\sigma^{2}}\exp\left(-\frac{A^{2}}{2\sigma^{2}}\right)\text{.}
\]
Enforcing a certain threshold $A_{0}$ gives a total probability 
\[
P_{0}\left(A_{0}\right)=\int_{0}^{A_{0}}\mathrm{d}A'\,p(A')=1-\mathrm{e}^{-\frac{A_{0}^{2}}{2\sigma^{2}}}
\]
and its inverse
\[
A_{0}\left(P_{0}\right)=\sqrt{-2\sigma^{2}\ln\left(1-P_{0}\right)}\text{.}
\]
With this formula, we selected $b$ to be the $99.999\%$ threshold,
i.e., $b=A_{0}(0.99999)$.}{\small\par}

{\small{}For the two scenarios of curve extraction and imaging with
a Gaussian convolution, the parameter $\sigma$ of the Rayleigh distribution
remains to be determined. Starting with the derivation of speckle
statistics \cite{goodman2007speckle}, it can be computed to yield
\[
\sigma=\frac{1}{\sqrt{2\left|M\right|}}
\]
for curve extraction and 
\[
\sigma=\frac{1}{\sqrt{8\pi}\sigma_{\text{Gauss}}}
\]
for imaging, when assuming unit magnitude phasors (as present in \eqref{eq:WeightsCurve}
and \eqref{eq:WeightsImaging}) with completely random phases. }{\small\par}

\subsubsection{{\small{}Phase unwrapping of the cross-spectrum}}

{\small{}While alternate approaches to extract the phase from the
cross-spectrum deal with the phase wrapping problem differently (e.g.\cite{Haniff:91}),
for our scenario phases of the obtained cross-spectrum need to be
unwrapped in the $\Delta t$ axis in order to obtain good results.
However, in 1D phase unwrapping, single outliers, i.e., a random phase
for single specific $\Delta t$, can tremendously degrade results
for all following values. We therefore slightly modified the standard
1D-phase unwrapping approach:}{\small\par}

{\small{}We assume the initial $D$ phases to be free of wrapping.
Afterwards we moved through the data set from small $\Delta t$ to
larger $\Delta t$. We computed the sum of all absolute values of
the phase differences to the preceding $D$ phases for each phase
value that is to be determined. We increased or decreased the respective
phase value by $2\pi$ as long as this difference sum kept decreasing.
Then me move to the next $\Delta t$. This procedure was done once
for curve extraction and repeated for each $x$ and $y$ value for
imaging. In our scenario, we used $D=10$.}{\small\par}

\subsubsection{{\small{}Obtaining the phase from the cross spectrum}}

{\small{}Instead of using the approaches described by \eqref{eq:PhaseDifferences}
or \eqref{eq:IntPhaseDifferences}, we formulate the recovery of the
phases from the cross-spectrum as a linear least squares problem.
This also served as basis for many of the previously demonstrated
approaches in astronomy. Assuming the phase of the cross-spectrum
is unwrapped in its $\Delta t$ axis, we can assum}e that 
\begin{multline}
\sum_{t,\Delta t}w^{2}(\vec{x};t,\Delta t)\\
\times\left|\phi(\vec{x},t+\Delta t)-\phi(\vec{x},t)-\left\langle C_{U}(\vec{x};t,\Delta t)\right\rangle \right|^{2}\label{eq:CS-Minimization}
\end{multline}
minimizes for{\small{} the desired phase $\phi\left(\vec{x},t\right)$,
if the weights $w$ represent the quality of the cross-spectrum for
the respective parameters $\vec{x}$, $t$, and $\Delta t$. However,
the solution to this linear least square problem is not unique: since
the cross-spectrum only contains phase differences, there will be
different solutions for different initial values $\phi\left(\vec{x},t=0\right)$.
In addition, all weights might be $0$ for one specific $t$ which
represents a gap in reliable data. Both problems in the evaluation
can be encountered by Tikhonov regularization with small parameters.
For the former case one should force the resulting phase for one time
point $t_{0}$ to be small; the corresponding regularization parameter
should not influence other results as long as it is chosen sufficiently
small to not introduce numerical errors. For the second problem, a
difference regularization can be introduced. Again, the parameter
can be small; it is only used is to obtain a unique solutions, in
case weights approach $0$. Consequently, the entire approach can
be formulated as a linear least squares problem, and solved by regularized
solving of the respective linear equation. }{\small\par}

{\small{}To solve the least squares problem \eqref{eq:CS-Minimization}
we first need to discretize it properly. To this end, we first create
the discrete vectors and matrices $\vec{c}_{\phi}$ corresponding
to the unwrapped cross-spectrum phase $\arg\left(\left\langle C_{U}\right\rangle \right)$,
$\vec{\phi}$ corresponding to the phases to be extracted, and a matrix
$S$ relating the two. Assume a given systematic phase $\vec{\phi}'$,
then the corresponding cross-spectrum $\vec{c}_{\phi}'$ would be
uniquely given by }{\scriptsize{}
\[
\vec{c}_{\phi}'=\left(\begin{array}{c}
\phi'_{0}-\phi'_{1}\\
\phi'_{0}-\phi'_{2}\\
\vdots\\
\phi'_{1}-\phi'_{2}\\
\vdots
\end{array}\right)=\underbrace{\left(\begin{array}{ccccc}
1 & -1 & 0 & \cdots & 0\\
1 & 0 & -1 & \cdots & 0\\
 &  & \vdots & \ddots\\
0 & 1 & -1 & \cdots & 0\\
 &  & \vdots
\end{array}\right)}_{S}\left(\begin{array}{c}
\phi'_{0}\\
\phi'_{1}\\
\vdots\\
\phi'_{N-1}
\end{array}\right)\text{.}
\]
}{\small{}We can write this as 
\[
\vec{c}_{\phi}'=S\vec{\phi}'\text{.}
\]
Given the actually measured cross-spectrum $\vec{c}_{\phi}$ and the
corresponding phase $\vec{\phi}$ to be computed and introducing the
diagonal weight matrix $W=\diag\left(w_{0},w_{1,}...,w_{N-1}\right)$
as computed by \eqref{eq:WeightsCurve} or \eqref{eq:WeightsImaging}
we can write the determination of $\phi$ as the regularized minimization
problem by 
\[
\left\Vert W\left(S\phi-c_{\phi}\right)\right\Vert ^{2}+\mu_{1}\left\Vert \Gamma_{1}\phi\right\Vert ^{2}+\mu_{2}\left\Vert \Gamma_{2}\phi\right\Vert ^{2}\to\min\text{.}
\]
The corresponding $\vec{\phi}$ is found by 
\[
\vec{\phi}=\left(S^{T}W^{2}S+\mu_{1}^{2}\Gamma_{1}^{T}\Gamma_{1}+\mu_{2}^{2}\Gamma_{2}^{T}\Gamma_{2}\right)^{-1}S^{T}W^{2}\vec{c}_{\phi}\text{,}
\]
which needs to be performed for each curve or each lateral pixel $(x,y)$
when doing phase imaging. In general, we chose 
\[
\Gamma_{1,ij}=\begin{cases}
1 & \text{for \ensuremath{i=j=4}}\\
0 & \text{otherwise}
\end{cases}
\]
and
\[
\Gamma_{2,ij}=\begin{cases}
1 & \text{if \ensuremath{i=j}\ensuremath{\neq0}}\\
-1 & \text{if \ensuremath{i=j+1}}\\
0 & \text{otherwise,}
\end{cases}
\]
with $\mu_{1}=0.02$ and $\mu_{2}=0.001$. }{\small\par}

\subsubsection{{\small{}Specifying the initial value}}

{\small{}The regularization term $\mu_{1}\left\Vert \Gamma_{1}\right\Vert ^{2}$
basically enforces the phase value corresponding to the 5th volume
to 0, thereby specifying the initial value. Since in both, simulation
and experiment, we only know that no (deliberate) phase change is
occurring in the frames $<5$, we can normalize the final result by
\[
\phi_{\text{res},i}=\phi_{i}-\frac{1}{N_{0}}\sum_{j=0}^{N_{0}-1}\phi_{j}\text{,}
\]
with $N_{0}=5$.}{\small\par}

\subsubsection{{\small{}Real-valuedness of the modulating cross-spectrum}}

{\small{}For the entire approach to work it remains to be shown, that
the modulation cross spectrum $\left\langle C_{H}\right\rangle $
is real-valued for small $\Delta t$. We assume the modulation cross-spectrum
is given according to \eqref{eq:Modulation} and \eqref{eq:ModulationAvgCrossSpectrum}
by }{\footnotesize{}
\begin{align*}
\left\langle C_{H}\right\rangle = & \left\langle H(t)H^{*}(t+\Delta t)\right\rangle \\
= & \left\langle \left(\sum_{i}A_{i}(t)\mathrm{e}^{\mathrm{i}\phi_{i}\left(t\right)}\right)\left(\sum_{j}A_{i}(t+\Delta t)\mathrm{e}^{\mathrm{i}\phi_{i}\left(t+\Delta t\right)}\right)^{*}\right\rangle \\
= & \Bigl\langle\sum_{i}A_{i}\left(t\right)A_{i}^{*}\left(t+\Delta t\right)\mathrm{e}^{\mathrm{i}\left(\phi_{i}\left(t\right)-\phi_{i}\left(t+\Delta t\right)\right)}\\
 & +\sum_{i,j,i\neq j}A_{i}\left(t\right)A_{j}^{*}\left(t+\Delta t\right)\mathrm{e}^{\mathrm{i}\left(\phi_{i}\left(t\right)-\phi_{j}\left(t+\Delta t\right)\right)}\Bigr\rangle\\
= & \left\langle \sum_{i}A_{i}\left(t\right)A_{i}^{*}\left(t+\Delta t\right)\mathrm{e}^{\mathrm{i}\left(\phi_{i}\left(t\right)-\phi_{i}\left(t+\Delta t\right)\right)}\right\rangle \\
 & +\left\langle \sum_{i,j,i\neq j}A_{i}\left(t\right)A_{j}^{*}\left(t+\Delta t\right)\mathrm{e}^{\mathrm{i}\left(\phi_{i}\left(t\right)-\phi_{j}\left(t+\Delta t\right)\right)}\right\rangle \text{.}
\end{align*}
}{\small{}Now, the second term is small compared to the first term
since $\phi_{i}$ and $\phi_{j}$ are statistically independent making
the sum and the average run over random phasors. The phases of the
first term can be approximated by $\phi_{i}(t)-\phi_{i}(t)+\left.\partial_{t'}\phi_{i}\left(t'\right)\right|_{t'=t}\Delta t=\left.\partial_{t'}\phi_{i}\left(t'\right)\right|_{t'=t}\Delta t$
and thus will be $0$ for small $\Delta t$ that are within good autocorrelation
of the modulating function $H$. Thus $\left\langle C_{H}\right\rangle $
is real for small $\Delta t$ to a good approximation.}{\small\par}

\subsection{{\small{}Phase evaluation by phase differences}}

{\small{}In the alternative approach we compute merely phase differences.
We can still formulate this by using the cross-spectrum of phase difference
$C_{U}^{Z_{1}-Z_{2}}$ given by \eqref{eq:CrossSpectrumPhaseDiff}.
For all values $t\ge t_{0}$, the phase is obtained by 
\[
\phi(t)=\arg\left\{ G_{\mathrm{Gauss}}*C_{U}^{Z_{1}-z_{2}}\left(x,y;t_{0},t\right)\right\} 
\]
for imaging and 
\[
\phi\left(t\right)=\arg\sum_{\left(x,y\right)\in M}C_{U}^{Z_{1}-Z_{2}}\left(x,y;t_{0},t\right)
\]
for curve evaluation. For this direct phase computation, the phase-wrapped
cross-spectrum was used. For curve extraction the phase was unwrapped
in $t$ afterwards.}{\small\par}
\end{document}